\documentclass[pra,aps,reprint,superscriptaddress,amsmath,longbibliography]{revtex4-1}
\usepackage{graphicx}
\usepackage{dsfont}
\usepackage{bm}
\usepackage[hidelinks]{hyperref}

\newcommand{\bra}[1]{\langle #1 | \,}
\newcommand{\ket}[1]{\, | #1 \rangle}

\newcommand{\hlf}{\mbox{$\frac{1}{2}$}}
\newcommand{\De}{\Delta}
\newcommand{\Dec}{\Delta_{\mathrm{c}}}
\newcommand{\gcc}{g_{\mathrm{c}}}
\newcommand{\Ga}{\Gamma}
\newcommand{\gac}{\gamma_{\mathrm{c}}}
\newcommand{\hrho}{\hat{\rho}}
\newcommand{\nth}{\bar{n}_{\mathrm{th}}}
\newcommand{\neff}{\bar{n}_{\mathrm{eff}}}

\newcommand{\hsig}{\hat{\sigma}}

\begin{document}

\title{Faithful state transfer between two-level systems 
via an actively cooled finite-temperature cavity}

\author{L\H{o}rinc S\'ark\'any}
\author{J\'ozsef Fort\'agh}
\affiliation{Physikalisches Institut, Eberhard Karls Universit\"at T\"ubingen, 
D-72076 T\"ubingen, Germany}

\author{David Petrosyan}
\affiliation{Physikalisches Institut, Eberhard Karls Universit\"at T\"ubingen, 
D-72076 T\"ubingen, Germany}
\affiliation{Institute of Electronic Structure and Laser, 
FORTH, GR-71110 Heraklion, Crete, Greece}

\date{\today}

\begin{abstract}
We consider state transfer between two qubits -- effective two-level 
systems represented by Rydberg atoms -- via a common mode of 
a microwave cavity at finite temperature. We find that when both qubits 
have the same coupling strength to the cavity field, at large enough detuning 
from the cavity mode frequency, quantum interference between the transition 
paths makes the \textsc{swap} of the excitation between the qubits largely 
insensitive to the number of thermal photons in the cavity. 
When, however, the coupling strengths are different, the photon 
number-dependent differential Stark shift of the transition frequencies 
precludes efficient transfer. 
Nevertheless, using an auxiliary cooling system to continuously extract 
the cavity photons, we can still achieve a high-fidelity state transfer 
between the qubits.  
\end{abstract}

\maketitle

\section{Introduction}

Quantum information protocols and quantum simulations with cold atomic 
systems extensively utilize strong dipole-dipole interaction between 
the laser-excited Rydberg states of atoms 
\cite{Lukin2001, Saffman2010, Saffman2016, Petrosyan2017, Schauss1455, Labuhn2016, Bernien2017,Lienhard2017}. 
Atoms are routinely trapped and manipulated near the surfaces 
of superconducting atom chips \cite{Fortagh2007,Bernon2013}, 
and can couple to on-chip microwave planar resonators 
\cite{Verdu2009,Hattermann2017}. 
Atom chips are typically cooled by liquid helium to the temperature 
of $T\simeq 4$~K. 
At such temperatures, the black body radiation is sufficiently suppressed 
and does not detrimentally affect the Rydberg state lifetime 
\cite{Beterov2009,Saffman2016,Mack2015}. Rydberg qubits then exhibit 
long coherence times on atom chips \cite{Avigliano2014,Teixeira2015}. 
This should be contrasted with superconducting qubits that require
chips at $T\lesssim 100$~mK temperature to operate
\cite{Wallraff2004, Chiorescu2004, You2011}. 

Huge electric dipole transitions between the Rydberg states of atoms
allow their strong coupling to microwave planar waveguides \cite{Hogan2012}
and resonators \cite{Petrosyan2009}. The cavity field can then serve
as a quantum bus to mediate long distance interactions and quantum gates 
between the Rydberg qubits \cite{Petrosyan2008,Pritchard2014}. 
At $T\simeq 4$~K, however, the microwave cavity or waveguide has large 
population of thermal photons. This would preclude long-distance quantum 
gates and coherent state transfer, unless inherently temperature-resistant 
schemes are used \cite{Sarkany2015, Xiang2017, Vermersch2017}. 

In this paper we consider a \textsc{swap} operation between a pair of Rydberg 
qubits exchanging a virtual photon via a thermally populated microwave cavity.
We show that high-fidelity state or excitation transfer is only possible 
if the coupling strengths of both atoms to the cavity mode is the same. 
In the general case of unequal couplings, the photon number-dependent 
Stark shifts of the atomic transitions precludes the state transfer. 
In order to overcome this problem, we propose to continuously extract thermal 
photons from the cavity mode using a separate, laser-driven ensemble of 
cooling atoms. This breaks the symmetry between the cavity field population 
from the thermal environment and depopulation due to the photon decay 
and extraction, greatly reducing the effective mode temperature. 
We find that despite the increased rate of photon decay from the cavity,
high-fidelity \textsc{swap} operation between the Rydberg qubits is 
feasible with a realistic experimental setup. 

The paper is organized as follows. In Sec.~\ref{sec:system} we describe
the system and show that the excitation transfer between two atoms 
in non-equivalent positions of a thermal cavity is suppressed. 
In Sec.~\ref{sec:cavcool} we present a method to reduce the number of
the thermal photons of the cavity by their continuous extraction with 
an optically pumped atomic ensemble. In Sec.~\ref{sec:cavcooltrnans}
we show that this leads to increased transfer probability between the 
two atoms, followed by conclusions in Sec.~\ref{sec:conclud}

\section{The compound system}
\label{sec:system}

\subsection{Mathematical formalism}

\begin{figure}[t]
\centerline{\includegraphics[width=8.7cm]{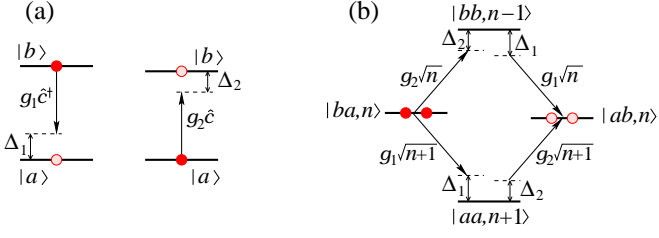}}
\caption{
Schematics of the system. 
(a) Two atoms with levels $\ket{a},\ket{b}$ are coupled to the cavity field 
$\hat{c}$ with the coupling strengths $g_{\mathrm{1}}$ and $g_{\mathrm{2}}$ 
and detunings $\Delta_{\mathrm{1}}$ and $\Delta_{\mathrm{2}}$, respectively. 
(b) Transitions between states $\ket{ba,n}$ and $\ket{ab,n}$ containing
$n$ cavity photons can occur via two paths involving intermediate 
non-resonant states $\ket{bb,n-1}$ and $\ket{aa,n+1}$. }
\label{fig:ALscheme}
\end{figure}

We consider a compound system consisting of two two-level (Rydberg) 
atoms $i=1,2$ coupled to a common mode of a microwave resonator, 
as sketched in Fig.~\ref{fig:ALscheme}(a). The system is thus
equivalent to the Jaynes--Cummings model for two atoms interacting
with a cavity field \cite{PLDP2007,Haroche2006}.  
In the frame rotating with the frequency of the cavity mode $\omega$,
the Hamiltonian is given by
\begin{eqnarray}
\mathcal{H}/\hbar &=& \sum_{i=1,2} 
\big[ \hlf \De_i ( \hsig_{bb}^{(i)} - \hsig_{aa}^{(i)} ) 
+ g_i ( \hat{c} \hsig_{ba}^{(i)} + \hsig_{ab}^{(i)} \hat{c}^{\dag}) \big]. 
\end{eqnarray}
Here $\De_{i} \equiv \omega_{ba}^{(i)} - \omega$ is the detuning of atom 
$i$ on the transition $\ket{a} \leftrightarrow \ket{b}$ from the cavity 
mode frequency, $\hsig_{\mu \nu}^{(i)} \equiv \ket{\mu}_i\bra{\nu}$ 
($\mu,\nu=a,b$) are atomic operators, and $g_i$ is the coupling strength 
of atom $i$ to the cavity mode described by the photon creation 
$\hat{c}^{\dag}$ and annihilation $\hat{c}$ operators. The coupling 
strength $g_i = (\wp_{ab}/\hbar) \varepsilon_{\mathrm{c}} u(\mathbf{r}_i)$
is proportional to the dipole matrix element $\wp_{ab}$ of the atom
on the transition $\ket{a} \to \ket{b}$, field per photon 
$\varepsilon_{\mathrm{c}}$ within the cavity volume, and the cavity mode
function $u(\mathbf{r}_i)$ at the position $\mathbf{r}_i$ of the atom. 
Since atoms 1 and 2 may be at non-equivalent positions, $g_1$ and $g_2$
are in general different. We also assume that the detunings $\De_1$ and
$\De_2$ can be individually controlled by, e.g., spatially inhomogeneous
electric (dc Stark) or magnetic (Zeeman) fields or by a focused 
non-resonant laser inducing an ac Stark shift of one of the Rydberg 
states \cite{Leseleuc2017}.

Several relaxation processes are affecting the system. 
We assume that both Rydberg states $\ket{a}$ and $\ket{b}$ 
decay with approximately the same rate $\Gamma$  to some lower state(s) 
$\ket{s}$ which are decoupled from the cavity field. These decay processes 
are described by the Liouvillians \cite{PLDP2007,Haroche2006}
\begin{eqnarray}
\mathcal{L}_{\mathrm{a}_i} \hrho &=& 
\hlf \Gamma (2 \hsig_{sa}^{(i)} \hrho \hsig_{as}^{(i)} - \hsig_{aa}^{(i)} \hrho - \hrho \hsig_{aa}^{(i)}) 
\nonumber \\ &  &
+ \hlf \Gamma (2 \hsig_{sb}^{(i)} \hrho \hsig_{bs}^{(i)} - \hsig_{bb}^{(i)} \hrho - \hrho \hsig_{bb}^{(i)}) 
\end{eqnarray}  
acting onto the density operator $\hrho$ of the total system. 
The relaxation of the cavity field toward the thermal equilibrium
with rate $\kappa$ is described by
\begin{eqnarray}
\mathcal{L}_{\mathrm{c}} \hrho &=&
\hlf \kappa (1+\nth) (2 \hat{c} \hrho  \hat{c}^{\dag}
-  \hat{c}^{\dag} \hat{c}\hrho - \hrho \hat{c}^{\dag} \hat{c} ) 
\nonumber \\ &  &
+ \hlf \kappa \nth (2 \hat{c}^{\dag} \hrho \hat{c} - \hat{c} \hat{c}^{\dag} \hrho
- \hrho  \hat{c} \hat{c}^{\dag}), \label{eq:Lc} 
\end{eqnarray} 
where $\nth = (e^{\hbar \omega/k_{\mathrm{B}}T} -1)^{-1}$ is the mean number 
of thermal photons in the cavity mode at temperature $T$
\cite{PLDP2007,Haroche2006}.  

The density operator of the total system obeys the master equation
\begin{equation}
\partial_t \hrho = -\frac{i}{\hbar} [\mathcal{H},\hrho] + 
\mathcal{L}_{\mathrm{a}_1}\hrho + \mathcal{L}_{\mathrm{a}_2}\hrho
 + \mathcal{L}_{\mathrm{c}}\hrho. \label{eq:MasterEq}
\end{equation}
We solve numerically the equations for the density matrix of the system
whose Hilbert space is a tensor product space of two three-state atoms, 
$\{\ket{a},\ket{b}, \ket{s} \}_{i=1,2}$, and the cavity field with the photon 
number states $\{\ket{n}\}$ truncated at sufficiently large $n \leq 100$. 

\subsection{Adiabatic elimination of the cavity mode}

We are interested in the state or excitation transfer between 
the atoms 1 and 2 using the cavity mode as a quantum bus. 
Consider the states $\ket{ba,n}$ and $\ket{ab,n}$ with either 
one or the other atom excited, while the cavity mode contains 
$n$ photons, see Fig.~\ref{fig:ALscheme}(b).
There are two transition paths between these states via the 
intermediate states $\ket{aa,n+1}$ and $\ket{bb,n-1}$ involving 
a photon addition to or subtraction from the cavity mode.
In order to minimize the effects of relaxation and thermalization of 
the cavity mode during the transfer, the atoms  
should exchange virtual cavity photons. We therefore choose 
the atomic detunings to be similar and large enough, 
$\Delta_{\mathrm{1,2}} \gg |\De_1 - \De_2|, g_{\mathrm{1,2}} \sqrt{n_{\mathrm{max}}}$,
where $n_{\mathrm{max}}$ is the maximal number of photons that can be in the 
cavity with appreciable probability; typically, $n_{\mathrm{max}}$ can be taken 
as $10 \times \nth$, and we recall that at thermal equilibrium the 
probability distribution of the cavity photon number is 
\begin{equation}
P_n = \frac{\nth^n}{(1+\nth)^{n+1}}.
\end{equation} 

Using the perturbation theory, we adiabatically eliminate 
the nonresonant intermediate states $\ket{aa,n+1}$ and $\ket{bb,n-1}$.
We then obtain that the energies of states $\ket{ba,n}$ and $\ket{ab,n}$  
are Stark shifted by the cavity field as 
\begin{subequations}
\begin{eqnarray}
E_{ba,n} & \simeq & \hlf \delta + \frac{g_1^2(n+1)}{\De_1 +S(n+1)} 
- \frac{g_2^2 n}{\De_2+Sn}, \\
E_{ab,n} & \simeq & - \hlf \delta + \frac{g_2^2(n+1)}{\De_2 +S(n+1)} 
- \frac{g_1^2 n}{\De_1 +Sn},
\end{eqnarray}
\end{subequations}
where $\delta \equiv \De_1 - \De_2$ and 
$S \equiv \frac{g_1^2}{\De_1} + \frac{g_2^2}{\De_2}$. 
Simultaneously, the transition amplitude 
between the states $\ket{ba,n}$ and $\ket{ab,n}$ is given by
\begin{eqnarray}
G(n) &\simeq &\frac{g_1 g_2 (n+1)}{2}
\left[\frac{1}{\De_1 +S(n+1)} +  \frac{1}{\De_2 +S(n+1)}\right] 
\nonumber \\ & & 
- \frac{g_1 g_2 n}{2} \left[\frac{1}{\De_1 +Sn} +  \frac{1}{\De_2 +Sn}\right] ,
\end{eqnarray}
where the first and the second terms correspond to the amplitudes of 
transitions via the states $\ket{aa,n+1}$ and $\ket{bb,n-1}$, respectively.

\subsection{Transfer probability}

\begin{figure}[t]
\centerline{\includegraphics[width=7.5cm]{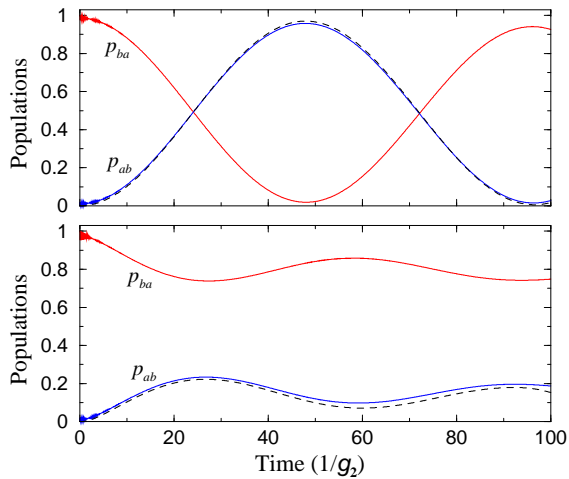}}
\caption{
Oscillations between the states $\ket{ba}$ (red solid line) and 
$\ket{ab}$ (blue solid line) in a cavity, as obtained from the
numerical simulations of the Master equation (\ref{eq:MasterEq}).
The parameters are $\De_1 = \De_2 = 30 g_2$, 
$\kappa = 10^{-3} g_2$, $\nth = 5$, 
$\Gamma = 3\times 10^{-4} g_2$,  
while $g_1 = g_2$ (upper panel) and $g_1 = 1.4 g_2$ (lower panel).
The dashed black line results from the approximate expression 
(\ref{eq:p0bosc}).}
\label{fig:oscil}
\end{figure}

States $\ket{ba,n}$ and $\ket{ab,n}$ have the energy difference
$\delta E(n) \equiv E_{ba,n} - E_{ab,n}$
and are coupled with rate $G(n)$.
Since we assumed $\De_1 \approx \De_2 = \Delta \gg \delta$, we can cast 
the energy difference and transition rate as
\begin{eqnarray}
\delta E(n) & \simeq & \delta + \frac{g_1^2 - g_2^2}{\Delta} (2n+1) 
, \label{eq:deEn} \\
G(n) &\simeq & \frac{g_1 g_2}{\Delta} \left( 1-\frac{\delta}{2\Delta} 
-  \frac{g_1^2 + g_2^2}{\Delta^2} (2n+1) \right), \label{eq:Geffn}
\end{eqnarray}
where we neglected terms of the order of $\delta \frac{g^2}{\Delta^2}$, 
$\frac{g^4}{\Delta^3}$ and higher. 
We thus see that both $\delta E(n)$ and $G(n)$ depend on the cavity 
photon number $n$. If the coupling strengths $g_1$ and $g_2$ are 
the same, the differential Stark shifts of levels $\ket{ba,n}$ and 
$\ket{ab,n}$ becomes $n$ independent. Then, by choosing $\delta =0$,
we obtain that both atoms have the same transition frequency, 
$\delta E(n) = 0 \, \forall \,n$. Simultaneously, due to interference 
between the two transition paths, $G(n)$ only weakly depends on the 
cavity photon number $n$, leading to resonant exchange of excitation 
between the atoms. In Fig.~\ref{fig:oscil} upper panel, we show the 
oscillations between the states $\ket{ba}$ and $\ket{ab}$ of the two
atoms with equal couplings $g_1=g_2$ to the cavity which has a large 
mean thermal photon number $\nth$ corresponding to a broad photon 
number distribution $P_n$.

When, however, the coupling strengths $g_1$ and $g_2$ are different,
the energy difference $\delta E(n)$ is photon number dependent. Then, 
in a thermal cavity, $\nth > 0$, with a broad distribution $P_n$ of photon 
numbers, $\delta E(n)$ cannot be made to vanish for all $n$. In fact
the difference between $\delta E(n)$ for various $n$ is of the order
of the transition rate $G$, which suppresses the amplitude of oscillations
between the states $\ket{ba}$ and $\ket{ab}$, as shown in 
Fig.~\ref{fig:oscil} lower panel. 

We can derive an approximate expression for the transfer probability 
between the states $\ket{ba}$ and $\ket{ab}$ as follows.
Assume that we start in state $\ket{ba,n}$ at time $t=0$. Without relaxations, 
the probability for the system to be in state $\ket{ab,n}$ would be given by
\begin{equation*}
p_{ab,n}(t) = \left| \frac{G(n)}{\bar{G}(n)} \right|^2 
\sin^2 \left[\bar{G}(n) t \right],
\end{equation*}
where $\bar{G}(n) = \sqrt{|G(n)|^2 + \frac{1}{4} |\delta E(n)|^2}$ is the 
generalized Rabi frequency for the oscillations between $\ket{ba,n}$ 
and $\ket{ab,n}$.
Relaxations will result in exponential damping of the transfer probability.
The decay of states $\ket{a}$ and $\ket{b}$ of each atom with rate $\Ga$ 
leads to multiplication by $e^{- \Ga t}$. 
We neglect for now the cavity relaxation and will consider its effect later. 
Thus the transfer probability from state $\ket{ba,n}$ to state $\ket{ab,n}$ 
is given by
\begin{equation}
p_{ab,n}(t) = e^{- 2 \Ga t} 
\left| \frac{G(n)}{\bar{G}(n)} \right|^2 \sin^2 \left[ \bar{G}(n) t \right],
\end{equation}
If initially there is an equilibrium photon number distribution $P_n$ in the 
cavity, the total probability of transfer between $\ket{ba}$ and $\ket{ab}$ 
is given by
\begin{equation}
p_{ab}(t) = \sum_n P_n p_{ab,n}(t) , \label{eq:p0bosc} 
\end{equation}
This expression approximates well the exact dynamics of the system
for small $\kappa \ll G$ as verified in Fig.~\ref{fig:oscil}.

\section{Reducing the number of thermal photons} 
\label{sec:cavcool}

The thermal photons in the cavity thus preclude efficient state 
transfer between the atoms when their couplings to the cavity mode have, 
in general, different strength, $g_1 \neq g_2$. 
We now outline a method to reduce the number of photons in the cavity, 
which will significantly increase the transfer probability. 
We will use an ensemble of stationary trapped atoms extracting photons 
from the cavity by continuous optical pumping, attaining thereby an 
equilibrium with a smaller mean number of photons. 
We note a conceptually similar approach \cite{Raimond2001,Nogues1999,Bernu2008} 
to extract thermal photons from a microwave cavity by sending across 
it a sequence of atoms prepared in the lower Rydberg state, which is 
typically done in the transient regime to achieve a nearly empty cavity 
until it equilibrates with the thermal environment.

\subsection{Cavity cooling by photon extraction}

\begin{figure}[t]
\centerline{\includegraphics[width=7.5cm]{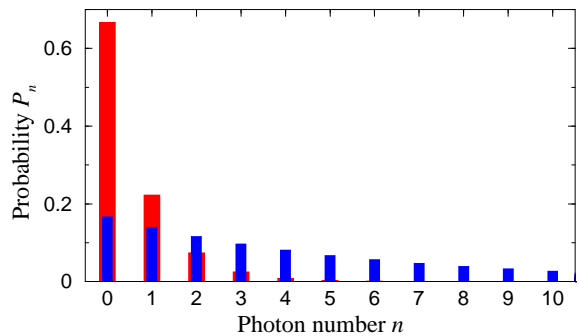}}
\caption{
Equilibrium photon number distribution $P_n$ in the cavity with
$\nth =5$ (blue narrow bars) and with continuous photon extraction with 
rate $\gac = 9 \kappa$ (red thicker bars) leading to $\neff = 0.5$.}
\label{fig:pncool}
\end{figure}

Our strategy to cool the cavity is to continuously extract the photons 
from it with a rate $\gac \gg \kappa$. This process is described
by the Liouvillian 
\begin{eqnarray}
\mathcal{L}_{\mathrm{c}}^{\prime} \hrho &=&
\hlf \gac (2 \hat{c} \hrho  \hat{c}^{\dag}
-  \hat{c}^{\dag} \hat{c}\hrho - \hrho \hat{c}^{\dag} \hat{c} ) , 
\end{eqnarray} 
which should be added to $\mathcal{L}_{\mathrm{c}} \hrho$ in Eq.~(\ref{eq:Lc}).
The photon extraction thus breaks the balance between the usual 
photon decay to and addition from the thermal reservoir. 
It follows from the Master equation for the cavity field that
the photon number probabilities obey the equation 
\[
\partial_t P_n = d(n+1) P_{n+1} + a n P_{n-1} - d n P_n - a (n+1) P_n ,
\]
where $a= \kappa \nth$ and $d = \kappa (\nth + 1) + \gac$. 
In the steady state, $\partial_t P_n = 0$, we have the detailed balance 
$d (n+1) P_{n+1} = a (n+1) P_{n}$ [and $d n P_{n} = a n P_{n-1}$] for any
transition $n \leftrightarrow n \pm 1$. This leads to $P_n = (a/d)^n P_0$,
and upon normalization of the probability distribution $P_n$, we obtain
the usual expression 
\begin{equation}
P_n = \frac{\neff^n}{(1+\neff)^{n+1}} ,
\end{equation} 
corresponding to the equilibrium with the effective thermal photon
number $\neff = \frac{\nth}{1 + \gac/\kappa}$ reduced from $\nth$
by a factor of $(1 + \gac/\kappa)$. 
In Fig.~\ref{fig:pncool} we show the equilibrium photon number distribution 
in a cavity at a temperature corresponding to large $\nth$. Upon 
continuous photon extraction with rate $\gac \gg \kappa$, we obtain
a thermal distribution, with a much smaller effective photon number $\neff$, 
which is highly peaked around $n=0$. 

\subsection{Photon extraction by optical pumping}

\begin{figure}[t]
\centerline{\includegraphics[width=2.5cm]{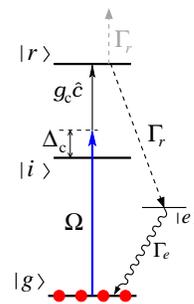}}
\caption{
Schematics of the cooling setup. 
Atoms in the ground state $\ket{g}$ are excited to the Rydberg state
$\ket{r}$ by a two-photon transition via intermediate nonresonant 
state $\ket{i}$. The resonant two-photon transition involves a
photon absorption from the laser field acting on the transition 
$\ket{g}\to\ket{i}$ with Rabi frequency $\Omega$ and detuning 
$\Dec \gg \Omega$, and a photon absorption from the cavity 
field coupled with strength $\gcc \ll \Dec$ to the Rydberg 
transition $\ket{i}\to\ket{r}$ detuned by $-\Dec$. 
State $\ket{r}$ decays with rate $\Gamma_r$ either by ionization 
or by an induced cascade to state $\ket{g}$ via intermediate state 
$\ket{e}$ having large decay rate $\Gamma_s$.}
\label{fig:coolsch}
\end{figure}

To extract the photons from the thermal cavity, we envision a system 
depicted in Fig.~\ref{fig:coolsch}. An ensemble of $N_\mathrm{c}$ ``cooling''
atoms in the ground state $\ket{g}$ are trapped near the cavity antinode. 
A laser field acts on the transition from state $\ket{g}$ to a Rydberg 
state $\ket{i}$ with the Rabi frequency $\Omega$ and large detuning 
$\Dec \gg \Omega$. Each atom is coupled to the cavity field $\hat{c}$
on the Rydberg transition $\ket{i}\to\ket{r}$ with strength 
$\gcc \ll \Dec$. Upon adiabatic elimination of the nonresonant
intermediate state $\ket{i}$, we obtain an effective Rabi frequency 
$\Omega_{n}^{(2)} = \Omega \gcc \sqrt{n}/\Dec$ for the two-photon 
transition $\ket{g,n} \to \ket{r,n-1}$ which involves absorption 
of a laser photon and a cavity photon. 

We can write the equations for the amplitudes $A_{g,n}$ and $A_{r,n-1}$ 
of states $\ket{g,n}$ and $\ket{r,n-1}$ as
\begin{subequations}
\begin{eqnarray}
\partial_t A_{g,n} &=& -i \Omega_{n}^{(2)} A_{r,n-1} , \\
\partial_t A_{r,n-1} &=& - \hlf \Gamma_r A_{r,n-1}  - i \Omega_{n}^{(2)} A_{g,n} ,
\end{eqnarray}
\end{subequations}
where $\Gamma_r$ is the population decay rate of the Rydberg state $\ket{r}$
and we assume the two-photon resonance. Assuming $\Gamma_r \gg \Omega_{n}^{(2)}$
(see below), we can set $\partial_t A_{r,n-1}=0$ and obtain the incoherent 
transition rate $\hlf R_n = \frac{|\Omega_{n}^{(2)}|^2}{\Gamma_r/2}$ from 
$\ket{g,n}$ to $\ket{r,n-1}$. With $N_\mathrm{c}$ cooling atoms, we can 
then identify the photon extraction rate via $\gac n = N_\mathrm{c} R_n$, 
leading to 
\begin{equation}
\gac = N_\mathrm{c} \frac{4 \Omega^2 \gcc^2}{\Dec^2 \Gamma_r} .
\end{equation}
Note that contributions of individual atoms add incoherently to the total 
extraction rate $\gac$ and the possible variation of the coupling strength
$\gcc$ for different atoms can be absorbed into redefinition of the atom 
number $N_\mathrm{c}$.
In the above analysis, we have also neglected the (Rydberg blockade) 
interactions between the atoms in state $\ket{r}$. This approximation 
is justified for moderate number of photons $n$ and large enough 
$\Gamma_r \gg \Omega_{n}^{(2)}$, such that the probability of having 
simultaneously two or more atoms in state $\ket{r}$ is small. 
 
Typically, Rydberg states have slow population decay rates. Larger decay
rate $\Gamma_r$ can be achieved by laser-induced ionization of state $\ket{r}$, 
which, however, will result in continuous depletion of the number $N_\mathrm{c}$ 
of cooling atoms. A better alternative is to use an auxiliary laser 
of Rabi frequency $\Omega_{r}$ to resonantly couple the Rydberg state 
$\ket{r}$ to a lower excited state $\ket{e}$ having large decay rate 
$\Gamma_e \gg \Omega_{r}$ back to the ground state $\ket{g}$, 
as shown in Fig.~\ref{fig:coolsch}. This will induce a cascade
from $\ket{r}$ to $\ket{g}$ with sufficiently rapid rate 
$\Gamma_r \simeq \frac{4 \Omega_{r}^2}{\Gamma_e}$. Thus, each cooling
cycle $\ket{g,n} \to \ket{r,n-1} \to \ket{g,n-1}$ will extract with
rate $\gac$ a cavity photon, while the number of cooling atoms will 
be preserved. For $\gac \gg \kappa \nth$, we can then approach a 
cavity vacuum by optically pumping out thermal photons.  

\section{State transfer via cooled cavity}
\label{sec:cavcooltrnans}

\begin{figure}[t]
\centerline{\includegraphics[width=7.5cm]{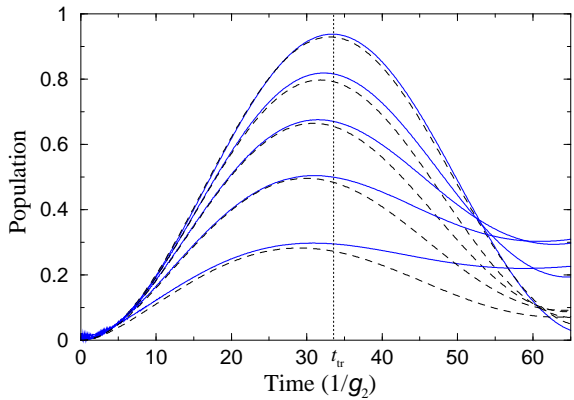}}
\caption{
Time dependence of population $p_{ab}$ of state $\ket{ab}$ in the presence 
of continuous photon extraction with different rates $\gac$, as obtained 
from exact numerical simulations of Eq.~(\ref{eq:MasterEq}) (blue solid lines) 
and from the approximate Eq.~(\ref{eq:p0bosc}) (black dashed lines), 
for the initial state $\ket{ba}$. Dotted vertical line indicates
time $t_{\mathrm{tr}}=\frac{\pi}{2 G(0)}$ when the population 
of $\ket{ab,0}$ is peaked. 
The parameters are $g_1 = 1.4 g_2$, $\De = 30 g_2$,
$\delta = \frac{g_2^2 -g_1^2}{\Delta}$, $\Gamma = 3\times 10^{-4} g_2$,
$\kappa = 10^{-2} g_2$, $\nth = 5$, while $\gac = (0, 1.5, 4, 9, 49) \kappa$ 
corresponding to the mean photon numbers $\neff = (5, 2, 1, 0.5, 0.1)$ 
for the graphs with progressively larger amplitudes. }
\label{fig:oscilcool}
\end{figure}

In Fig.~\ref{fig:oscilcool} we demonstrate significant enhancement of
the amplitude of oscillations between the initial $\ket{ba}$ and the
target $\ket{ab}$ states in the presence of rapid extraction of thermal
photons from the cavity. Since the probability distribution of the cavity
photons is now highly peaked at $n=0$, we set the frequency mismatch 
$\delta = \frac{g_2^2 -g_1^2}{\Delta}$ to satisfy the resonant
condition $\delta E(0) = 0$ in Eq.~(\ref{eq:deEn}).

\subsection{Optimizing the transfer fidelity}

From Eq.~(\ref{eq:Geffn}) we have for the transition rate
$G(0) \simeq \frac{g_1 g_2}{\Delta} \left( 1-\frac{\delta}{2\Delta} 
-  \frac{g_1^2 + g_2^2}{\Delta^2} \right)$. Our aim is to transfer 
the population of state $\ket{ba}$ to state $\ket{ab}$.
Using Eq.~(\ref{eq:p0bosc}), we can estimate the lower bound for 
the transfer probability as being determined mainly by the $n=0$ term, 
\begin{eqnarray}
p_{ab}(t) >  P_0 p_{ab,0}(t) &=& \frac{1}{1+\neff} e^{- 2 \Ga t} 
\nonumber \\ & & 
\times \hlf \{ 1 - \cos[2G(0)t] e^{-\kappa_{\mathrm{eff}} t} \} ,  \quad 
\end{eqnarray}
where we included the effective damping rate 
$\kappa_{\mathrm{eff}} = (\kappa+\gac) \frac{g_1^2 +g_2^2}{2\Delta^2}$ 
of the oscillation amplitude, which can be intuitively understood as follows: 
During the transfer, states $\ket{ba,0}$ and $\ket{ab,0}$ have small admixture, 
$\sim \frac{g_{\mathrm{1,2}}}{\Delta}$, of state $\ket{aa,1}$ containing 
the additional exchange photon which is damped with rate $(\kappa + \gac)$.
The transfer probability is peaked at time $t_{\mathrm{tr}}=\frac{\pi}{2 G(0)}$
when $\cos[2G(0)t_{\mathrm{tr}}] = -1$. Assuming $\neff \ll 1$ and
$(\kappa_{\mathrm{eff}},\Ga) t_{\mathrm{tr}} \ll 1$, the transfer 
fidelity is then 
\begin{eqnarray*}
F & \equiv & P_0 p_{ab,0}(t_{\mathrm{tr}}) 
\nonumber \\ 
& \gtrsim  & 
\left[ 1 - \frac{\kappa \nth}{\kappa + \gac} \right]
\left[ 1 - \pi \frac{\Gamma\Delta }{g_1 g_2} \right]
\left[ 1 - \pi \frac{\kappa + \gac}{4	\Delta} \frac{g^2}{g_1 g_2} \right] ,
\label{eq:Flb}
\end{eqnarray*}
where $g^2 = \hlf(g_1^2 +g_2^2)$.
Although the right-hand side of this expression underestimates the maximal 
fidelity, we can still use it to optimize the parameters of the system. 
Thus, the transfer fidelity is reduced by three factors, and we therefore
require that each of them be small:
\begin{itemize}
\item[(i)] $\gac + \kappa \gg \kappa \nth$, i.e. the cooling rate $\gac$ should 
be sufficiently large to have the mean photon number small, $\neff \ll 1$;
\item[(ii)] $\frac{g_1 g_2}{\Delta} \gg \Gamma$, i.e., the transition
rate $G(0)$ should be large enough to have small probability of the  
atomic decay during the transfer.   
\item[(iii)] $\Delta \gg \gac + \kappa$, i.e., the cavity field should be 
strongly detuned to have small photon population and therefore decay during 
the transfer. 
\end{itemize}

With $\gac \gg \kappa$, the total reduction of the fidelity, or infidelity, 
can be estimated as
\begin{equation}
1-F \simeq \frac{\kappa \nth}{\gac} + 
\pi \frac{\Gamma \Delta }{g_1 g_2} + 
\pi \frac{\gac}{4\Delta} \frac{g^2}{g_1 g_2} .
\end{equation}
We can minimize this expression with respect to $\Delta$, with 
the other parameters fixed, obtaining 
\begin{equation}
1-F \simeq \frac{\kappa \nth}{\gac} 
+ \frac{\pi g}{g_1 g_2} \sqrt{ \Gamma \gac} ,
\end{equation}
for $\Delta = g \sqrt{\frac{\gac}{4\Gamma} }$.
Next, we minimize the resulting infidelity with respect to 
$\gac$, finally obtaining 
\begin{equation}
\min(1-F) \leq 3 (\kappa \nth)^{1/3} 
\left( \frac{\pi g \sqrt{\Gamma}}{2 g_1 g_2} \right)^{2/3} 
\end{equation}
for $\gac = \left( \frac{2 \kappa \nth g_1 g_2}{\pi g \sqrt{\Gamma}} \right)^{2/3}$.

\subsection{Experimental considerations}

We assume the following realistic parameters of the system:
The cavity mode resonant frequency is $\omega = 2 \pi \times 5\:$GHz, 
and its quality factor is $Q = 10^5$ leading to the decay rate 
$\kappa \simeq 300\:$kHz.
The Rydberg atom--cavity coupling rates are $g_1 =  2 \pi \times 5\:$MHz 
and similar for $g_2$.
The decay rates of the atoms are $\Ga = 10\:$kHz. With $\nth \sim 10$ 
at cryogenic environment, 
we need to choose the cooling rate $\gac \gtrsim 15 \:$MHz,
which can be achieved with $N_{\mathrm{c}} \sim 1000$ cooling 
atoms, with $\Omega \simeq \gcc =  2 \pi \times 0.1\:$MHz, 
$\Dec = 10 \Omega$ and $\Gamma_r = 1\:$MHz.
We then choose the detunings $\De_1 \simeq 30  g_1$ and 
$\De_2 = \De_1 + \frac{g_1^2 - g_2^2}{\Delta}$ for the resonant 
transfer of the excitation between the atoms.
The resulting fidelity is $F \gtrsim 0.95$ which we verified numerically.
To turn off the transfer, one of the atoms can be strongly detuned by 
$|\De_1 - \De_2| \gg G(0) \simeq \frac{g_1 g_2}{\Delta}$, which can 
be achieved by, e.g., Stark shifting the resonance with a focused laser
beam \cite{Leseleuc2017}.

\section{Conclusions}
\label{sec:conclud}

We have elaborated the conditions for coherent state transfer between two 
two-level systems through a thermal microwave cavity. We have demonstrated 
that by actively cooling a cavity mode by continuously removing photons 
with a laser-driven ensemble of atoms, high-fidelity \textsc{swap} 
operation between pairs of spatially separated Rydberg-atom qubits 
is possible in state-of-the-art experimental systems 
\cite{Hattermann2017,Avigliano2014,Teixeira2015,Hogan2012}. 
The $\sqrt{\mbox{\textsc{swap}}}$ is a universal entangling quantum 
gate \cite{PLDP2007}, which can also be realized by the present scheme. 

Trapped ground-state atoms have good coherence properties and can serve 
as reliable qubits. The atoms can be excited on demand by focused 
lasers to the Rydberg states for realizing short distance quantum 
communication and quantum logic gates. Our results will thus pave the way 
for the realization of scalable quantum information processing with cold 
atoms trapped on the integrated superconducting atom chips.

\begin{acknowledgments}
We acknowledge the financial support of the 
DFG Schwerpunktsprogramm Giant interactions in Rydberg Systems (GiRyd SPP 1929).
\end{acknowledgments}
 
\bibliography{Literature}

%
%
%

%
%

\end{document}